\documentclass[twocolumn,english,prl]{revtex4-1}
\usepackage[T1]{fontenc}
\usepackage[latin9]{inputenc}
\usepackage[letterpaper]{geometry}
\usepackage{color}
\usepackage{relsize}
\usepackage{graphicx}
\usepackage{amssymb}

\makeatletter

\DeclareRobustCommand{\greektext}{%
  \fontencoding{LGR}\selectfont\def\encodingdefault{LGR}}
\DeclareRobustCommand{\textgreek}[1]{\leavevmode{\greektext #1}}
\DeclareFontEncoding{LGR}{}{}

\DeclareRobustCommand{\lyxmathsym}[1]{\ifmmode\begingroup\def\b@ld{bold}
  \def\rmorbf##1{\ifx\math@version\b@ld\textbf{##1}\else\textrm{##1}\fi}
  \mathchoice{\hbox{\rmorbf{#1}}}{\hbox{\rmorbf{#1}}}
  {\hbox{\smaller[2]\rmorbf{#1}}}{\hbox{\smaller[3]\rmorbf{#1}}}
  \endgroup\else#1\fi}

\@ifundefined{textcolor}{}
{%
 \definecolor{BLACK}{gray}{0}
 \definecolor{WHITE}{gray}{1}
 \definecolor{RED}{rgb}{1,0,0}
 \definecolor{GREEN}{rgb}{0,1,0}
 \definecolor{BLUE}{rgb}{0,0,1}
 \definecolor{CYAN}{cmyk}{1,0,0,0}
 \definecolor{MAGENTA}{cmyk}{0,1,0,0}
 \definecolor{YELLOW}{cmyk}{0,0,1,0}
 }

\makeatother

\usepackage{babel}

\begin{document}

\title{EPR entanglement strategies in two-well BEC }

\author{Q. Y. He$^{1}$, M. D. Reid$^{1,2}$, C. Gross$^{2}$, M. Oberthaler$^{2}$,
and P. D. Drummond$^{1,2}$}

\email{pdrummond@swin.edu.au}

\affiliation{$^{1}$ARC Centre of Excellence for Quantum-Atom Optics, Centre for
Atom Optics and Ultrafast Spectroscopy, Swinburne University of Technology,
Melbourne 3122, Australia\\
$^{2}$Kirchhoff-Institut für Physik, Universität Heidelberg, Im
Neuenheimer Feld 227, 69120 Heidelberg, Germany}
\begin{abstract}
Criteria suitable for measuring entanglement between two different
potential wells in a Bose-Einstein condensation (BEC) are evaluated.
We show how to generate the required entanglement, utilizing either
an adiabatic two-mode or dynamic four-mode interaction strategy, with
techniques that take advantage of s-wave scattering interactions to
provide the nonlinear coupling. The dynamic entanglement method results
in an entanglement signature with spatially separated detectors, as
in the Einstein-Podolsky-Rosen (EPR) paradox.
\end{abstract}
\maketitle
One of the most important questions in modern physics is the problem
of macroscopic spatial entanglement, which directly impinges on the
nature of reality. Here we analyse how rapid advances in Bose-Einstein
condensation (BEC) in ultra-cold atoms can help to resolve this issue.
Recently, the observation of spin-squeezing has shown that measurement
beyond the standard quantum limit is achievable \cite{Esteve2008,Gross2010,Riedel2010}.
Spin squeezing is known to demonstrate entanglement between atoms
\cite{Sorenson}, but not which subsystems have been entangled. An
important step forward beyond this would be to realise quantum entanglement
in the Einstein-Podolsky-Rosen (EPR) sense; that is, having two spatially
separated condensates entangled with each other \cite{Hines}. This
is an important milestone towards future experiments involving entanglement
of macroscopic mass distributions, thereby demonstrating quantum Schroedinger
cat type superpositions of distinct mass distributions.

In this Letter, we analyse some achievable entangled quantum states
using a two-well BEC, and the measurable criteria that can be used
to signify entanglement. The types of quantum state considered include
number anti-correlated states prepared using adiabatic passage, as
well as dynamically prepared spin-squeezed states. In particular,
we focus on spin-entanglement, as a particularly useful route for
achieving measurable EPR entanglement, without requiring atomic local
oscillators. We note that spin orientation is easily coupled to magnetic
forces to allow superpositions of different mass distributions, once
spin entanglement is present. We consider different types of spin
entanglement criteria, and analyze which quantum states these are
sensitive to. 

\begin{figure}[h]
\begin{centering}
\includegraphics[width=0.85\columnwidth]{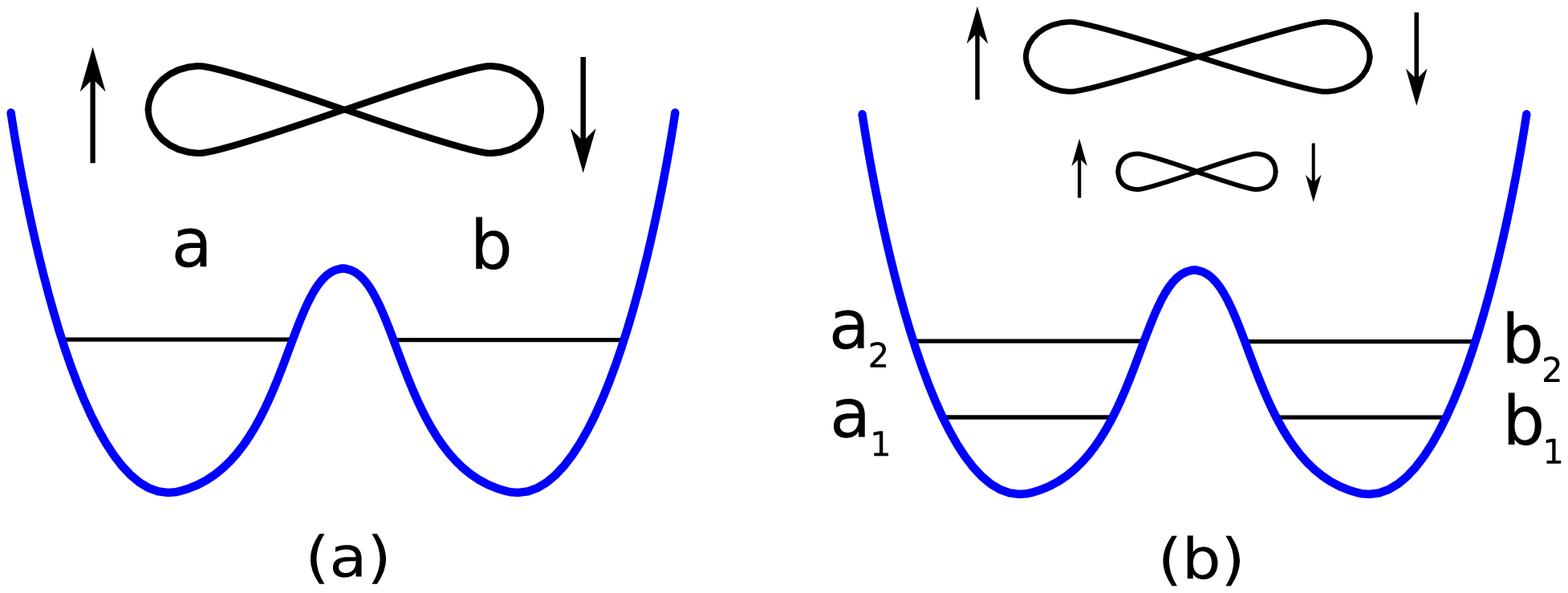}
\par\end{centering}

\caption{(a) Two internal modes $a_{,\ }b$ with  spatial entanglement; (b)
two pairs of modes $a_{1,\ }a_{2}$ and $b_{1,\ }b_{2}$ are entangled.
\label{fig:Double-well}}

\end{figure}

We show that existing experimental techniques appear capable of generating
spatial entanglement, with relatively minor changes. There are several
possible routes available. Our most significant conclusion is that
the criterion used to measure entanglement must be chosen carefully.
Not all measures of entanglement are equivalent, and there is an important
question as to what one regards as the fundamental subsystems, ie,
particles or modes. The appropriate choice of measure depends on the
entangled state, how it is prepared, and what type of detection is
technologically feasible. To demonstrate and analyse this need to
adapt the criterion to the state, we choose here to analyse two and
four mode models of a BEC, indicated schematically in Fig. \ref{fig:Double-well},
where $a_{1}$, $a_{2}$ are operators for two internal states at
$A$ and $b_{1},\ b_{2}$ are operators for two internal states at
$B$. 

In the limit of tight confinement and small numbers of atoms, this
type of system can be treated using a simple coupled mode effective
Hamiltonian, of form:\begin{equation}
\hat{H}/\hbar=\kappa\sum_{i}\hat{a}_{i}^{\dagger}\hat{b}_{i}+\frac{1}{2}\left[\sum_{ij}g_{ij}\hat{a}_{i}^{\dagger}\hat{a}_{j}^{\dagger}\hat{a}_{j}\hat{a}_{i}\right]+\left\{ \hat{a}_{i}\leftrightarrow\hat{b}_{i}\right\} \ .\label{eq:Hamiltonian}\end{equation}
Here $\kappa$ is the inter-well tunneling rate between wells, while
$g_{ij}$ is the intra-well interaction matrix between the different
spin components.

\paragraph{Adiabatic preparation: }

We first consider two-mode states having a single spin orientation,
with number correlations established using adiabatic passage in the
ground state. This makes them practical to prepare following earlier
experimental approaches \cite{Esteve2008,Ketterle_Interferometer},
as shown in Fig. \ref{fig:Double-well}(a). A recent multi-mode analysis
shows that effects of other spatial modes may be relatively small
\cite{Ferris}. In a two-mode analysis, we assume that $a_{1}$ and
$b_{1}$ have been prepared in the many-body ground state of Eq (\ref{eq:Hamiltonian})
with a fixed number of atoms $N$, while the second pair of spin states
$a_{2}$ and $b_{2}$ remain in the vacuum state, so that we can write
$a\equiv a_{1}$ and $b\equiv b_{1}$. In these cases there is only
one nuclear spin orientation, and there is existing experimental data
on phase coherence and number correlations \cite{Esteve2008,Ketterle_Interferometer},
with $10dB$ relative number squeezing being maximally indicated.
A number of previous analyses have used entropic measures specific
to pure states to study entanglement. These signatures cannot be readily
measured, and are not applicable to realistic mixed states that are
typically created in the laboratory.

However, one generally demonstrate spatial entanglement between the
two wells $a$ and $b$ using the non-Hermitian operator product criterion
of Hillery and Zubairy (HZ) \cite{hillzub}. This is also related
to a recently developed continuous-variable Bell inequality criterion
\cite{cvbell2}. A sufficient entanglement criterion between $A$
and $B$ is the operator product measure: \begin{equation}
|\langle a^{\dagger}b\rangle|^{2}>\langle a^{\dagger}ab^{\dagger}b\rangle\,.\label{eq:HZ}\end{equation}
Interwell spin operators have already been measured in this environment.
These are defined as: $\hat{J}_{AB}^{X}=\left(\hat{a}^{\dagger}\hat{b}+\hat{a}\hat{b}^{\dagger}\right)/2;\,\,\hat{J}_{AB}^{Y}=\left(\hat{a}^{\dagger}\hat{b}-\hat{a}\hat{b}^{\dagger}\right)/(2i);\,\,\hat{J}_{AB}^{Z}=\left(\hat{a}^{\dagger}\hat{a}-\hat{b}^{\dagger}\hat{b}\right)/2;\,\hat{J}_{AB}^{\pm}=\hat{J}_{AB}^{X}\pm i\hat{J}_{AB}^{Y};\,\hat{N}_{AB}=\hat{N}_{A}+\hat{N}_{B}=\hat{a}^{\dagger}\hat{a}+\hat{b}^{\dagger}\hat{b}$. 

In spin language, the HZ criterion shows that spatial entanglement
is proved for any state when\begin{eqnarray}
E_{HZ} & = & \frac{\langle\Delta\hat{J}_{AB}^{+}\Delta\hat{J}_{AB}^{-}\rangle}{\langle\hat{N_{A}}\rangle}\nonumber \\
 & = & \frac{\frac{1}{4}\langle\left[N_{A}+N_{B}\right]^{2}\rangle-\langle\left[\hat{J}_{AB}^{Z}\right]^{2}\rangle}{|\langle J_{AB}^{X}\rangle|^{2}+|\langle J_{AB}^{Y}\rangle|^{2}}<1\,.\label{eq:HZ-entanglement}\end{eqnarray}
This has similarities to the spin squeezing criterion \cite{sorenson}
which has now been measured experimentally ~\cite{Esteve2008,Gross2010}.
However, a crucial difference is that the spatial entanglement criterion
(\ref{eq:HZ-entanglement}) involves an increased relative number
fluctuation, rather than the reduced relative number fluctuations
found with the spin-squeezing criterion. Theoretically, we find that
two-well entanglement exists in the ground state with the HZ criterion,
although suppressed for increasingly strong repulsive interactions.
This behaviour is also known from previous studies using an entropic
$\varepsilon(\rho)$ entanglement measure \cite{Hines,XieHai}. The
strongest theoretical entropic entanglement is found when all atom
numbers are equally represented in the superposition. We find that
the closest state to this `super-entangled' limit is obtained at a
critical value of $Ng/\kappa\simeq-2$. This attractive interaction
regime (as found in $^{41}K$ and $^{7}Li$ isotopes) gives rise to
a maximal spread in the distribution of numbers in each well. Maximum
entanglement results for this model have also been found \cite{XieHai}
for entropic entanglement measures. In our calculations, we account
for effects of finite temperatures by assuming a canonical ensemble
of $\hat{\rho}=\exp\left[-\hat{H}/k_{B}T\right]$, with an interwell
coupling of $\hbar\kappa/k_{B}=50nK$. Our result for the Hillery-Zubairy
operator product signature is graphed below. This shows that two-well
spatial entanglement is maximized for an attractive inter-atomic coupling,
and the effect is relatively robust against thermal excitations:

\begin{figure}[h]
\begin{centering}
\includegraphics[width=0.9\columnwidth]{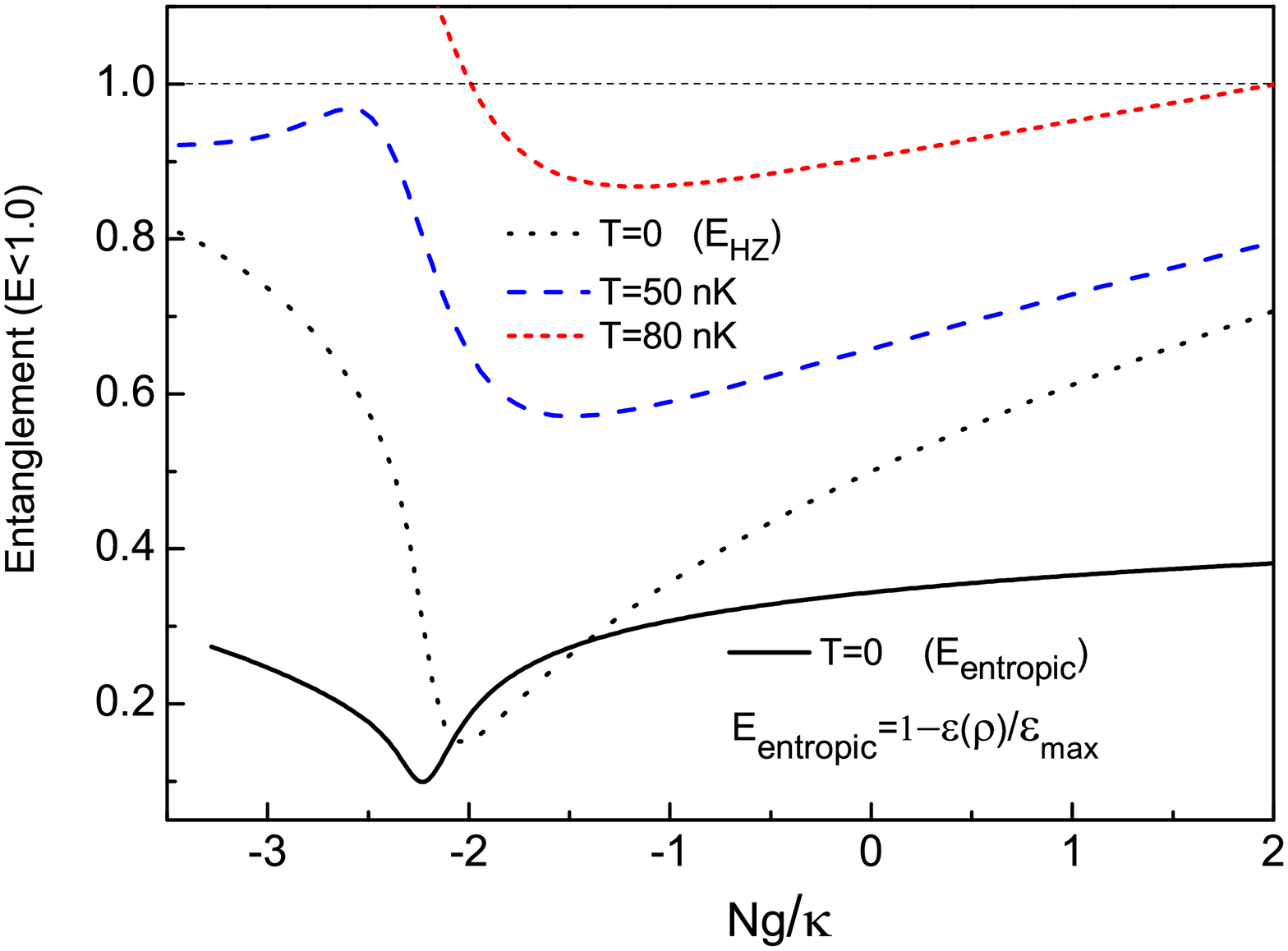}
\par\end{centering}

\centering{}\caption{Adiabatic entanglement with interactions in a two-well potential.
Dashed and dotted lines: HZ entanglement signature ($E_{HZ}<1$) at
$T=0K,\,50nK,\,80nK$ - lowest lines at lowest temperature; solid
line: entropic entanglement ($E_{entropic}=1-\varepsilon(\rho)/\epsilon_{max}<1$)
at $T=0K$. \label{fig:with adiabatic preparation}}

\end{figure}

\paragraph{Dynamic preparation: }

To proceed further, EPR entanglement as we define it requires using
measurements $O_{A}$ and $O_{B}$ that are individually defined either
at well $A$ or well $B$. Thus, entanglement is shown by performing
a set of simultaneous measurements on the spatially separated systems:
typically by measuring correlations $<O_{A}O_{B}>$ or $P(O_{A},O_{B})$.
This is necessary to justify Einstein's \textquotedbl{}no action at
a distance\textquotedbl{} assumption, that making one measurement
at $A$ cannot affect the outcome of another measurement at $B$.
One could achieve EPR entanglement with this criterion by making quadrature
amplitude measurements. That is, by expanding $a=X_{a}+iP_{a}$ and
$b=X_{b}+iP_{b}$, where $X_{a}$ etc are \textquotedbl{}quadrature
amplitudes\textquotedbl{}, so that the moment $<ab^{\dagger}>$ is
measured as four separate real correlations. Proposed methods for
measuring entanglement in BEC experiments include time-reversed dynamics
\cite{ZhangHelmersonSpinEnt}, and interference with side-modes of
a BEC moving through an optical lattice \cite{FerrisOlsen}. This
shows that, in principle, such a quadrature-based entanglement measurement
is not impossible. However, while feasible optically, this type of
measurement is nontrivial with ultra-cold atoms owing to interaction
induced phase fluctuations, and we propose a different strategy. 

To get good EPR measurements we consider instead the intra-well \textquotedbl{}spins\textquotedbl{}
$J^{X},\ J^{Y},$ and $J^{Z}$ at site $A$ and $B$.\textbf{\textcolor{red}{
}}$\ $This means having at least four modes in total. To prove EPR
entanglement using these measurements, one can define the spin measurements
at $A$ to be in terms of $a_{1}$ and $a_{2}$: $\hat{J}_{A}^{X}=\left(\hat{a_{1}}^{\dagger}\hat{a_{2}}+\hat{a_{2}}^{\dagger}\hat{a_{1}}\right)/2,$
$\hat{J}_{A}^{Y}=\left(\hat{a_{1}}^{\dagger}\hat{a_{2}}-\hat{a_{2}}^{\dagger}\hat{a_{1}}\right)/(2i),$
$\hat{J}_{A}^{Z}=\left(\hat{a_{1}}^{\dagger}\hat{a_{1}}-\hat{a_{2}}^{\dagger}\hat{a_{2}}\right)/2,$
$\hat{N}_{A}=\hat{a_{1}}^{\dagger}\hat{a_{1}}+\hat{a_{2}}^{\dagger}\hat{a_{2}}$;
also define raising and lowering operators as: $\hat{J}_{A}^{\pm}=\hat{J}_{A}^{X}\pm i\hat{J}_{A}^{Y}$,
and similar definition for site B. 

These are measurable locally using Rabi rotations and number measurements,
without local oscillators being required. The spin orientation measured
at each site can be selected independently to optimise the criterion
for the state used. One can then show EPR entanglement via spin measurements
using the spin version of the Heisenberg-product entanglement criterion
\cite{Bowen polarization ent-1} 

\begin{equation}
E_{product}=\frac{2\sqrt{\Delta^{2}\hat{J}_{AB}^{\theta\pm}\cdot\Delta^{2}\hat{J}_{AB}^{(\theta+\frac{\pi}{2})\pm}}}{|\langle J_{A}^{Y}\rangle|+|\langle J_{B}^{Y}\rangle|}<1\ ,\label{eq:product form}\end{equation}
or the sum criterion by Duan et al and Simon \cite{inseplur,FiberEntangle}

\begin{equation}
E_{sum}=\frac{\Delta^{2}\hat{J}_{AB}^{\theta\pm}+\Delta^{2}\hat{J}_{AB}^{(\theta+\frac{\pi}{2})\pm}}{|\langle J_{A}^{Y}\rangle|+|\langle J_{B}^{Y}\rangle|}<1\ ,\label{eq:SUM}\end{equation}
with general sum and difference spins $\hat{J}_{AB}^{\theta\pm}=\hat{J}_{A}^{\theta}\pm\hat{J}_{B}^{\theta}$,
and $J^{\theta}=cos(\theta)J^{Z}+sin(\theta)J^{X}$. Here the conjugate
Schwinger spin operators $J^{\theta}$ and $J^{\theta+\pi/2}$ obey
the uncertainty relation $\Delta^{2}J^{\theta}\Delta^{2}J^{\theta+\pi/2}\geq\frac{1}{4}|\langle J^{Y}\rangle|$.

In order to obtain ultra-cold atomic systems with four-mode entanglement,
we consider a dynamical approach to EPR entanglement which utilizes
phase as well as number correlations. This requires the BEC's to evolve
in time, in a similar way to successful EPR experiments in optical
fibres \cite{FiberEntangle,CRSD1987,fiber experiment }. This is very
different to the previous scheme, as the atom-atom interaction appears
explicitly as part of the time-evolution. The best entanglement is
obtained when the interaction between atoms of different spin is different
to the interaction between the atoms of the same spin. In Rubidium,
this either requires using a Feshbach resonance to break the symmetry,
or else separating the two spin components spatially as in the successful
fibre experiments \cite{fiber experiment } or in spin-squeezing atom-chip
experiments \cite{Riedel2010}. At a Feshbach resonance, for alkali
metals like Rubidium-87, the interactions between the different spin
orientations can be reduced compared to the self-interactions. This
allows an avenue for this type of entanglement with both the spin
orientations remaining \emph{in situ} in the same trap potential. 

To start with, we consider the conditions required to obtain the best
squeezing of Schwinger spin operators by optimizing the phase choice
$\theta$: $tg(2\theta)=2\langle J^{Z},\ J^{X}\rangle/(\Delta^{2}J^{Z}-\Delta^{2}J^{X})$.
Entanglement can be generated by the interference of two squeezed
states on a $50:50$ beam-splitter with a relative optical phase of
$\varphi=\pi/2$. This has also been achieved in optical experiments
\cite{FiberEntangle}, although not yet in BEC. 

\begin{figure}[h]
\begin{centering}
\includegraphics[width=0.9\columnwidth]{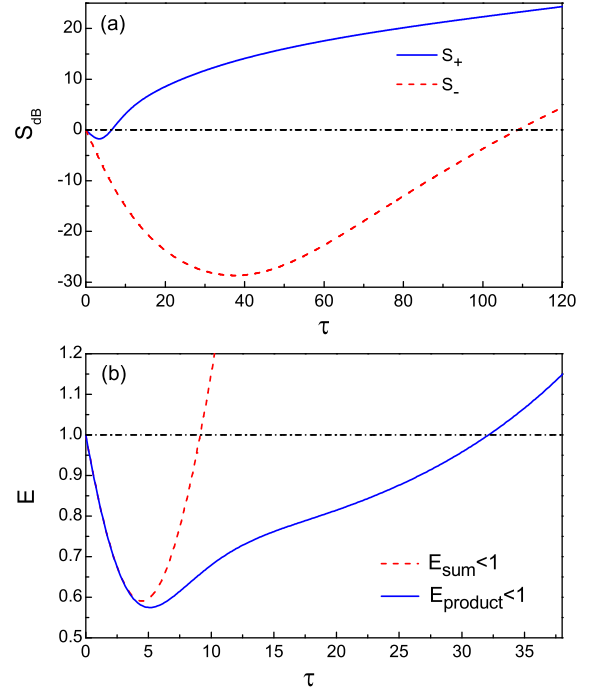}
\par\end{centering}

\centering{}\caption{(a) Squeezing of Schwinger spin operators $S_{dB}$: $S_{+}=10log_{10}\left[\Delta^{2}(J_{A}^{\theta}-J_{B}^{\theta})/n_{0}\right]$
(solid), $S_{-}=10log_{10}\left[\Delta^{2}(J_{A}^{\theta+\pi/2}+J_{B}^{\theta+\pi/2})\right]/n_{0}$
(dashed), and $n_{0}=\frac{1}{2}(|\langle J_{A}^{X}\rangle|+|\langle J_{B}^{Y}\rangle|)$
is shot noise (dotted). (b) Entanglement ($E_{product}$) based on
the criterion  (\ref{eq:product form}) by the solid curve and $E_{sum}$
in sum criterion (\ref{eq:SUM}) by the dashed curve. Here the parameters
correspond to $Rb$ atoms at magnetic field $B=9.131G$, with scattering
lengths $a_{11}=100.4a_{0}$, $a_{22}=95.5a_{0},$ and $a_{12}=80.8a_{0}$.
$a_{0}=53pm$. The coupling constant $g_{ij}\propto2\lyxmathsym{\textgreek{w}}_{\perp}a_{ij}$.
The number of $Rb$ atoms is $N_{A}=200$. $\tau=g_{11}N_{A}t$.\label{fig:with cross term}}

\end{figure}

Here we again take a four modes approach, explicitly assuming$a_{1},\ b_{1}$
and $a_{2},\ b_{2}$ that are initially in coherent states for simplicity,
i.e., assuming we have coherence between the wells. For simplicity,
we suppose that the initial state is then prepared in an overall four-mode
coherent state using a Rabi rotation. It is also possible to choose
a constrained total particle number, but we have used the simplest
model of coherence between the wells:\begin{equation}
|\psi>=|\alpha>_{a_{1}}|\alpha>_{b_{1}}|\alpha>_{a_{2}}|\alpha>_{b_{2}}\end{equation}

Next, we assume that the inter-well potential is increased so that
each well evolves independently. Finally, we decrease the inter-well
potential for a short time, so that it acts as a controllable, non-adiabatic
beam-splitter \cite{olsen}, to allow interference between the wells,
followed by independent spin measurements in each well. For dynamics,
we assume a simple two-spin evolution per well, which is exactly soluble.
We can treat this using either Schroedinger or Heisenberg equations
of motion. In the Heisenberg case, since the number of particles is
conserved in each mode, this has the solution:\begin{eqnarray}
\hat{a}_{i}\left(t\right) & = & \exp\left[-i\sum g_{ij}\hat{N}_{j}t\right]\hat{a}_{i}\left(0\right)\ ,\end{eqnarray}
where the couplings $g_{ij}$ are obtained from the known $Rb$ scattering
lengths at a Feshbach resonance.

\begin{figure}[h]
\begin{centering}
\includegraphics[width=0.9\columnwidth]{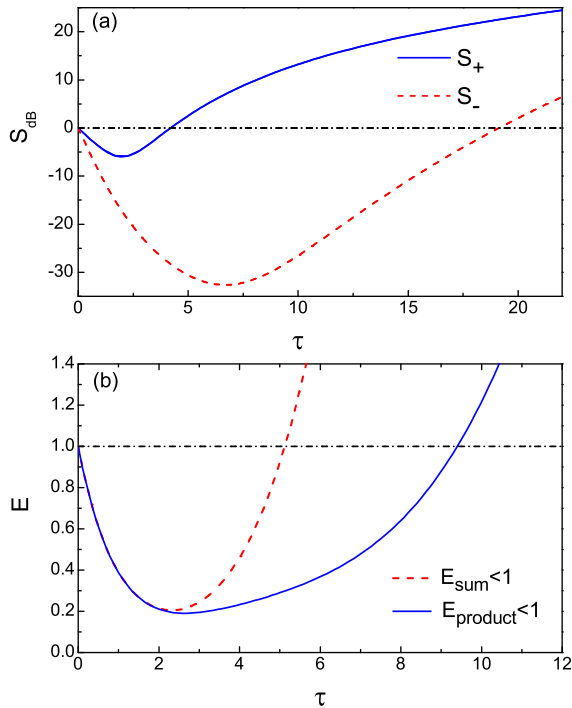}
\par\end{centering}

\centering{}\caption{Same as Fig. \ref{fig:with cross term} but assuming NO cross-couplings,
i.e., $g_{12}=0$. \label{fig:without cross term}}

\end{figure}

After dynamical evolution from an initial coherent state, we find
spin-squeezing in each well, prior to using the beam-splitter as shown
in Fig. \ref{fig:with cross term}(a). 

After using the beam-splitter, entanglement can be detected in principle
as $E<1$, as shown in Fig. \ref{fig:with cross term}(b). Note, Fig.
\ref{fig:without cross term} shows that assuming NO cross-couplings,
i.e., $g_{12}=0$ gives much better results than using the cross-couplings
obtained in a $Rb$ Feshbach resonance. As discussed in the adiabatic
approach, this would require spatially separated wells for the different
spin-orientations, or a different type of Feshbach resonance, in order
to eliminate cross-couplings.

In summary, we have shown two feasible techniques for measuring EPR-type
spatial BEC entanglement, using currently available double-well BEC
approaches combined with available atomic detection methods. The simplest
method employs an attractive ground-state adiabatic method, with a
single spin orientation. This requires an essentially nonlocal detection
strategy, in which the two BEC's are expanded and interfere with each
other. To obtain a spatially separated EPR entanglement strategy,
appropriate for investigating questions of local realism, we propose
a four-mode, dynamical strategy that employs two distinct spin orientations
in each spatial well.
\begin{acknowledgments}
We wish to thank the Humboldt Foundation, Heidelberg University, and
the Australian Research Council for funding via ACQAO COE and Discovery
grants, as well as useful discussions with Marcos de Oliveira. After
preparation of this manuscript, we learnt of a related calculation
by Nir Bar-Gill et. al, in arXiv:1009.2655.\end{acknowledgments}

\end{document}